# Hybrid integrated low noise linearly chirped Frequency Modulated Continuous Wave laser source based on self-injection to external cavity

LIWEI TANG,[1,2] HONGXAING JIA,[1,2] SHUAI SHAO,[1,2] SIGANG YANG,[1,2] HONGWEI CHEN,[1,2] AND MINGHUA CHEN,[1,2,*]

[1]*Department of Electronic Engineering, Tsinghua University, Beijing, 100084, China*
[2]*Beijing National Research Center for Information Science and Technology (BNRist), Beijing, 100084, China*
*\*chenmh@tsinghua.edu.cn*

**Abstract:** A low noise linearly frequency modulated continuous wave laser source with a wide frequency bandwidth is demonstrated. By two-dimensional thermal tuning, the laser source shows 42 GHz continuous frequency tuning with 49.86 Hz intrinsic linewidth under static operation. For dynamical FMCW, the laser source has 10.25 GHz frequency bandwidth at 100 Hz chirped frequency and 5.56 GHz at 1 kHz chirped frequency. With pre-distortion linearization, it can distinguish 3 m length difference at 45 km distance in the fibre length measured experiment, which demonstrate a potential for application in ultra-long fibre sensing and FMCW LiDAR.

## 1. INTRODUCTION

Frequency modulated laser source is the basic equipment of many detection systems such as the frequency modulated continuous wave (FMCW) light detection and ranging (LiDAR), pulse compression LiDAR, large-band linear chirped microwave generation and coherent optical frequency domain reflectometry (OFDR) [1, 2, 3, 4]. Compared to the conventional time-of-flight (ToF) LiDAR, the FMCW LiDAR is widely used in the autonomous vehicle (AV) and aerial photography due to the advantages of simultaneous measurement of velocity and location, the eye-safe continuous power and the high repeat frequency [5]. For FMCW coherent detection, the range resolution is relative to the frequency bandwidth and linearity of the frequency modulation. And a relatively longer chirped period provides abundant energy accumulation, which improves the signal to noise ratio (SNR) of the detected echo beat signal [6]. The time bandwidth product (TBWP) of the generated microwave can properly describe this characteristic. The narrow linewidth or low frequency noise of the LiDAR source provides a high coherent range, increasing the potential detectable distance.

However, for FMCW laser source, it is not easy to simultaneously satisfy narrow linewidth, high chirped linearity and large TBWP. For large tuning bandwidth, the famous Fourier-domain mode-locked (FDML) laser realized 100 nm tuning bandwidth with coherent continuous sweeping [7]. But it failed to achieve a low frequency noise because in the tens of kilometers fibre, the distributed environmental perturbation will induce the phase fluctuation of the stored longitudinal modes. And the chirped temporal duration is equal to the fibre time delay, longer temporal duration will induce serious instability, so it limits the TBWP to $1.5 \times 10^6$ [8, 9].

With the development of silicon photonics technology, it is a trend to reduce the linewidth by coupling the high quality (high-Q) micro resonator to semiconductor laser. By self-injection locking, the formed compound cavity extends the photon lifetime, which suppresses the amplified spontaneous emission (ASE) noise, narrowing the spectral linewidth and improving the coherent time. Coupling high-Q whispering gallery mode (WGM) microresonator to distributed feedback (DFB) laser, it achieves sub-100 Hz linewidth [10]. However, their scheme has no frequency modulation method under self-injection locked state.

In this paper, a FMCW laser source based on self-injection locked commercial distributed feedback (DFB) laser diode (LD) coupled with high-Q SiN micro-ring resonator (MRR) is demonstrated, which achieves large bandwidth with linear frequency modulation based on phase-MRR two-dimensional tuning method. Under static operation, 49.86 Hz intrinsic linewidth and 2.57 kHz integral linewidth with continuous tuning bandwidth up to 42 GHz is achieved. Under dynamic FMCW state, it achieves 10.25 GHz with chirped frequency of 100 Hz, and 5.56 GHz with chirped frequency of 1 kHz. The TBWP of the generated microwave waveform is up to $5.1 \times 10^7$. With pre-distortion linearization of the FMCW laser source, it can distinguish the length difference of 3 m at 45 km distance in the fiber length measurement experiment. With such a low noise, high linearity FMCW laser source, optical fibre sensing and LiDAR system will be further developed.

## 2. PRINCIPLE

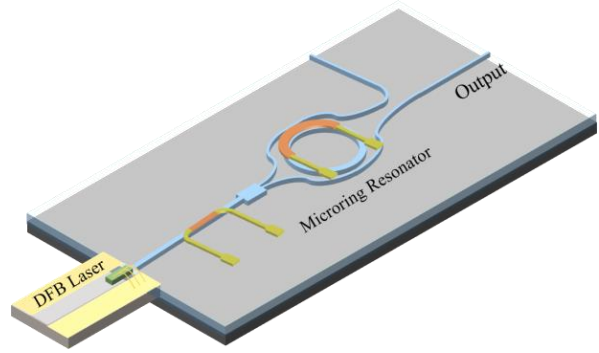

Fig. 1. Diagram of the proposed hybrid FMCM laser source. The DFB laser is coupled to an MRR where the blue line represents the waveguide, the red and yellow line represent the thermal detuning electrodes.

The proposed FMCW laser source consists of a DFB LD and an external MRR, as shown in Fig. 1. MRR is designed as the add-drop single ring type. The input port and drop port of the MRR are connected to the input waveguide by a multi-mode interferometer (MMI). The output port of the MRR is coupled to single mode fibre (SMF) as the output of the laser source. Another add port of MRR is left unconnected. Two thermal detuning electrodes are deposed on the silicon photonics chip. One of the electrodes is on the input waveguide before MMI for detuning the phase delay. Another is on the micro-ring waveguide for detuning the central resonant frequency. The optical feedback from MRR injects back to the DFB laser, enabling a great reduction of the linewidth.

The effect of the self-injection has been well studied. It can be divided by five regimes depending on the feedback ratio [11]. From low to high feedback ratio, the regimes are described as Regime I to Regime V. When the feedback efficiency is stronger than 10%, the DFB laser enters a Regime V which shows a stable single-longitudinal mode output with narrow linewidth. By detuning the MRR, the central resonant frequency is close to the DFB solitary frequency so that the feedback strength is improved. The DFB laser thus can be self-injection locked to the Regime V. The role of the MRR in the FMCW laser can be simplified into two parameters, viz. the full width at half maxima (FWHM) spectral linewidth at the central resonant frequency $\Lambda$, and the resonant frequency detuning from the solitary laser frequency $\omega_f$. The frequency-dependent complex reflection coefficient of the MRR can be expressed by the Lorenz function $\Gamma(\omega) = \frac{\Lambda}{\Lambda - i(\omega - \omega_f)}$. The dynamical response of the self-injected laser can be described by the well-known Lang-Kobayashi equation [12]:

$$(\omega - \omega_0)\tau = -C_{eff} \sin\left\{\omega\tau + arctan(\alpha) - \arctan(\frac{\omega - \omega_f}{\Lambda})\right\} \quad (1)$$

where

$$C_{eff} = \frac{\gamma\tau\Lambda\sqrt{1+\alpha^2}}{\sqrt{\Lambda^2 + (\omega - \omega_f)^2}} \quad (2)$$

The $C_{eff}$ represents the effective optical feedback strength. $\omega_0$ and $\omega$ represent the solitary frequency of the laser diode and the operating frequency of the output mode, respectively. $\tau$ is the external round-trip time, $\gamma$ is the feedback rate, $\alpha$ is the linewidth enhancement factor. $\psi = \omega\tau + arctan(\alpha)$ represents the external phase delay. Without loss of generality, $\psi$ can be represented by $\psi_0 + \omega\tau$. In our model, the $\psi_0$ and $\omega_f$ are detuned by the phase and MRR thermal electrodes, respectively.

By detuning the two electrodes, the laser can enter the self-injection locking state (Regime V). Fig 2 shows the output frequency $\omega$ detuning with respect to the MRR resonant frequency $\omega_f$ at the phase delay $\psi_0 = 2\pi/3$. The parameters select the commonly used value of the device [13]. As $\omega_f$ is tuned from low frequency to high frequency, the output frequency will follow the change and will jump to the locked area near the resonance frequency ($\omega_f = 0$ GHz). The output mode evolves to stable single-longitudinal mode state from multi-mode, mode-locked and chaotic state. Fig 2 also shows the hysteresis effect which is common in the self-injection laser [14]. When entering in the locked area by tuning $\omega_f$ from low frequency to high frequency, the $\omega_f$ can be tuned back lower for maximum frequency tuning range. The same process also occurs when tuning $\omega_f$ from high frequency to low frequency. The hysteresis effect does not affect the locking range. The frequency modulation relationship between $\omega_f$ and $\omega$ is constant.

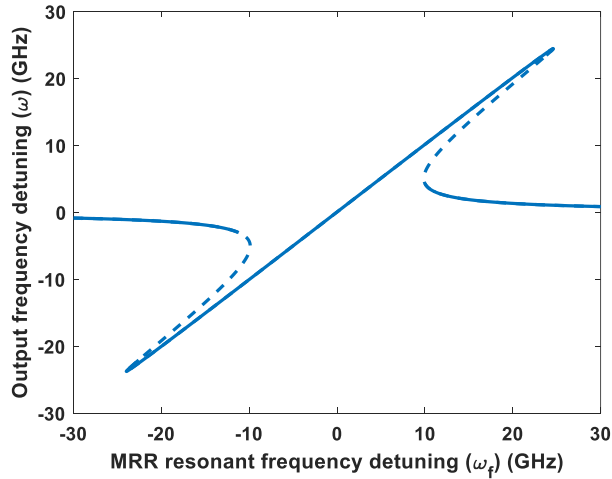

Fig. 2. Output frequency detuning with respect to the MRR resonance frequency.

The linear frequency modulated range in the locked state only occurs with the short external round-trip time. The external round-trip time should be less than 100 ps (equivalent to 2 cm cavity length). When applied to long round-trip time, the multi-valued and oscillating curve will be observed near the resonant frequency. By tuning $\omega_f$, $\omega$ will be detuned discretely which is mode hopping.

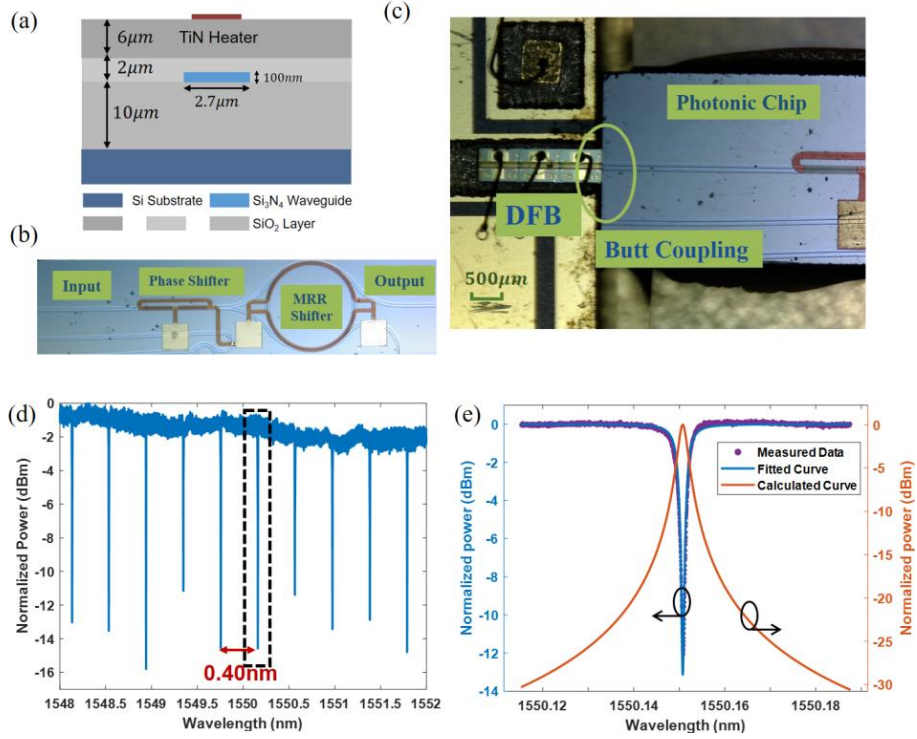

Fig. 3. (a) The cross section of the $Si_3N_4$ waveguide platform. (b) Microscope photo of the fabricated MRR. (c) Microscope photo of the hybrid laser. The DFB is butt coupled to MRR. (d) Transmission spectrum of MRR from input port to output port at the range of 1548 nm to 1552 nm. The FSR is 0.40 nm equivalent to 50 GHz. The area in black dashed line is enlarged in fig. 3(e). (e) Enlarged transmission spectrum at 1550 nm. The blue curve is the Lorentz-type fitting curve of the measured data. The red curve is reflected spectrum calculated by the fitted parameters.

The phase delay $\psi_0$ should be tuned to obtain a maximum tuning bandwidth. When tuning the $\omega_f$, the external cavity will also restrict the phase space of the self-injection laser. The phase delay $\psi_0$ needs to compensate for the phase detuning caused by the external cavity. Therefore, MRR's resonant frequency tuning needs to be accompanied by phase compensation tuning to achieve maximum FMCW bandwidth.

## 3. DESIGN AND FABRICATION

The MRR is fabricated based on the $Si_3N_4$ waveguide platform. It is a suitable platform for designing FMCW laser source. First, $Si_3N_4$ waveguide is transparent in the spectral range from 400 to 2350 nm [15], which can be freely applied to LiDAR source of different bands [16]. Second, the loss propagation waveguide enables a high-Q resonator fabrication, which is essential for narrowing the intrinsic linewidth of the LD. Third, the $Si_3N_4$ waveguide has high-aspect-ratio cross section [17]. The shape of the fundamental mode it supports matches well with the mode of the DFB laser. Therefore, the DFB laser can be butt coupled to the external cavity to obtain high coupling efficiency. The high coupling efficiency introduce high optical feedback ratio so that the DFB can enter the self-injection locked state. In addition, the $Si_3N_4$ material has relatively low thermo-optic coefficient (TOC) [18], which helps to improve the long-term thermal stability of the FMCW laser source.

The $Si_3N_4$ waveguide platform is used as our previous work [19]. As shown in Fig 3(a), the 100 nm-thick $Si_3N_4$ layer is grown on a 10 $\mu m$-thick thermal oxide $SiO_2$ layer. The $Si_3N_4$ layer is deposed by low pressure chemical vapor deposition (LPCVD) and waveguide is etched by

180 nm resolution lithography. The pattern mask is fabricated by reactive ion plasma etching (RIE). The etched $Si_3N_4$ waveguide is buried by a 2 $\mu m$-thick LPCVD $SiO_2$ and a 6 $\mu m$-thick plasma-enhanced chemical vapor deposition (PECVD) $SiO_2$ cladding layer. The $TiN$ thermal tuning electrode is deposed on the top of the waveguide.

The waveguide width is optimized to 2.7 $\mu m$ for low loss TE mode propagation. The full width at half maximum (FWHM) of the cold laser cavity of DFB laser diode is about 50 GHz [10]. To avoid multi-mode resonance, only a single feedback resonant frequency of the MRR is allowed to exist within the cold cavity bandwidth. So the diameter of the ring waveguide is designed to be 1.2 mm to achieve FSR of 50 GHz. The gap between ring waveguide and bus waveguide is optimized to be 1 $\mu m$ for high phase delay. The input waveguide is designed as a spot size converter (SSC) for mode field size matching with DFB. The fabricated MRR is demonstrated in Fig 3 (b).

The transmission spectrum from input port to output port is first measured. The advanced optical spectrum analyser (OSA, APEX 2062B) has a built-in wideband tunable laser source (TLS). Spectrum resolution of the OSA is of 0.08 pm (equivalent to 20 MHz), the spectrum bandwidth covers whole C-band. Fig 3 (d) demonstrates the transmission spectrum from 1548 nm to 1552 nm which shows FSR of 0.40 nm (equivalent to 50 GHz). The resonant frequency spectrum near 1550 nm is plot in Fig 3 (e). The raw measured data is drawn with purple dots, and fitted it by using standard add-drop MRR parameters [20]. The fitted curve is drawn in blue line and the normalized power is represented by the left axis. Using the fitted parameters, the reflected spectrum from drop-port is calculated which is drawn in red line and normalized power is represented by the right axis. The FWHM of the spectrum is 310 MHz and the Q value is calculated to be $6.19 \times 10^5$.

The DFB laser diode is commercially available. Different from the spatial coupling or vertical coupling [17, 21], the butt coupling is employed in this hybrid laser source, as shown in Fig 3 (c). The SSC is a straight waveguide with a gradually narrower width. The coupling loss is estimated to be less than -2.5 dB, which is able to introduce strong optical feedback to DFB laser diode (>10% feedback ratio). The output field from the integrated waveguide is guided in the single mode fibre (SMF) by a collimating lens and a lensed fibre. An optical isolator is placed between them to prevent external reflections. Thermistor and thermoelectric cooling (TEC) are also employed which are not shown in the Fig. 3(c).

## 4. EXPERIMENTAL RESULT

*4.1 Static Operation*

The test scheme is shown in Fig. 4(a). The OSA is used to monitor the output spectrum. By tuning the phase and MRR electrodes, the output mode evolves to self-injection locked state from multimode, chaotic state [22]. Fig. 4(b) shows the self-injection locked state at 1548.6 nm. It demonstrates a single-longitudinal mode with SMSR of 50 dB. Under self-injection locked state, tuning MRR and phase electrodes, it continuously shifts from 1548.631 nm to 1548.928 nm, equivalent to 42 GHz frequency bandwidth. The fibre coupled optical power is 0 dBm while the free-running power is 10 dBm. The power loss is mainly from the following four aspects: (1) Edge butt coupling loss between DFB and silicon photonic chip is -2.5 dB. (2) At the locked state, the oscillating mode is trapped in the MRR's resonant frequency. The optical power coupled from bus waveguide to ring waveguide is estimated to be about 3 dB. (3) The add port of MRR here is left unconnected, which has 3 dB loss. (4) Coupling from silicon photonic chip to SMF has 1 dB loss. In our design, the drop port and output port are not connected by an MMI in order to avoid interference effect and internal back reflections. In fact, studies have shown that the self-injection locked state has strong stability and anti-interference [23]. It has the potential to increase 3 dB fibre power by connected the two ports.

The frequency discriminating scheme is employed to measure the linewidth. The unbalanced Mach-Zehnder interferometer is used to obtain laser frequency noise spectrum at self-injection locked state [24]. As shown in Fig. 4(a), length difference of the two arms is

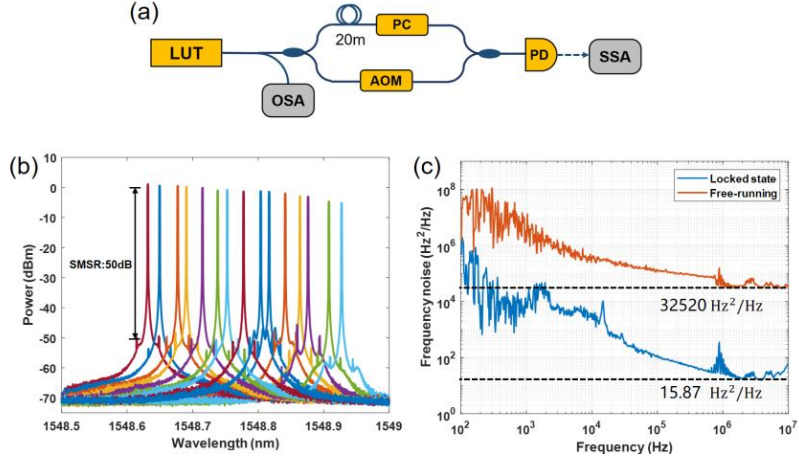

Fig. 4. (a) Experimental diagram of the test scheme. LUT: laser under test; PC: polarization controller; AOM: acousto-optic modulator; PD: photodiode; SSA: signal source analyser. (b) The optical spectrum under continuous frequency tuning. By tuning the phase and MRR electrode, the wavelength is detuned from 1548.631 nm to 1548.928 nm. The side-mode suppression ratio (SMSR) is 50 dB. (c) Frequency noise spectrum from 100 Hz to 10 MHz at static operation. The red and blue line are frequency noise spectrum of free-running and self-injection locked operation, respectively. White frequency noise of free-running and self-injection locked state are marked out in the figure.

20 m. AOM is placed at the shorter arm for heterodyne interference. The output signal is envelope detected by PD and is then fed into the SSA (E5052B, Keysight). The unbalanced MZI has sub-coherence time delay, the fluctuation of the laser frequency is converted to the amplitude fluctuations of the beat note. The frequency noise and laser intrinsic linewidth can be identified by analysing the phase noise spectral density (PNSD). This advanced measurement setup is similar to the more widely used delayed self-heterodyne interferometer (DSHI). However, in the measurement of ultra-narrow linewidth, the DSHI scheme loses its accuracy and the measurement of frequency noise is more reasonable, which has been analysed in detail [25, 26].

The SSA obtains the single-sideband power spectral density $S_{\Delta\varphi}(f)$ of the delayed laser phase noise where $\Delta\varphi = \varphi(t) - \varphi(t - \tau)$, $\tau$ is the time delay between two arms. The instantaneous laser frequency noise can be derived by [25]:

$$S_\nu(f) = \frac{f^2}{4[\sin(\pi f \tau)]^2} S_{\Delta\varphi}(f) \tag{3}$$

Frequency noise spectrum $S_\nu(f)$ of free-running and self-injection locked state from 100 Hz to 10 MHz are plotted in Fig. 4(c) respectively. At the high frequency range, the white noise is dominated which mainly caused by the random spontaneous emission and carrier fluctuations. This is so called "quantum-noise" processes that determines the Lorentzian-like shape. The intrinsic linewidth is directly referred by:

$$\Delta\nu_{intrinsic} = \pi S_\nu^0 \tag{4}$$

where $S_\nu^0$ is the single-sided power spectral density of the white frequency noise. In Fig. 4(c), the $S_\nu^0$ of free-running and self-injection locked state is read to be 32520 $Hz^2/Hz$ and 15.87 $Hz^2/Hz$ respectively. The intrinsic linewidth is calculated to be 102.16 kHz and 49.86 Hz. At locked state, the linewidth is supressed 3000 times. The ultra-low intrinsic linewidth has important applications in high speed communications. This improvement comes from the optical feedback of the high-Q external cavity. It provides a long photon lifetime and a pure spectral linewidth.

In coherent detection applications such as FMCW LiDAR and OFDR fibre sensing, the low frequency noise is dominant. The low frequency contributes to the beat note spectrum expansion near central frequency. The integrated linewidth is FWHM of the optical field power spectral density which only determined by the low frequency range. Integra linewidth can be solved by:

$$\int_{\Delta v_{int}}^{\infty} \frac{S_v(f)}{f^2} df = \frac{1}{\pi} rad^2 \tag{5}$$

where $S_v(f)/f^2$ represents the laser phase noise, $\Delta v_{int}$ represents the effective integral linewidth. It is approximated to be the low Fourier frequency which is the value of phase noise integration equals unity. In Fig. 4(c), the $\Delta v_{int}$ of free-running and self-injection locked state is solved to be 69.98 kHz and 2.57 kHz, respectively. The integral linewidth is more applicable in metrology and sensing. Compared with suppressing spontaneous emission to narrowing the intrinsic linewidth, the low frequency noise has more operable methods to suppress. The electrical feedback has been effectively reduce the frequency noise within its response bandwidth.

It is noted that the PNSD is measured at a static point in the locked state. Fig. 4(b) demonstrates the continuously frequency tuning for 42 GHz. When the wavelength is red shift from 1548.8 nm, the output power is gradually dropped. It is the state before mode-hopping where the locked position of MRR has shift. However, within the whole 42 GHz tunable band, the PNSD is maintained a low frequency noise as Fig. 4(c) shows. The intrinsic linewidth and integral linewidth are not observed significant change. And the single oscillating longitudinal mode as well as SMSR are also maintained in the whole tunable band. In self-injection dynamical system, the operation state switching is abrupt and the locked Regime V has hysteresis effect. Once enters the locked state, the narrow linewidth, single oscillating mode will maintain until switching to the next state.

*4.2 Linearly chirped FMCW*

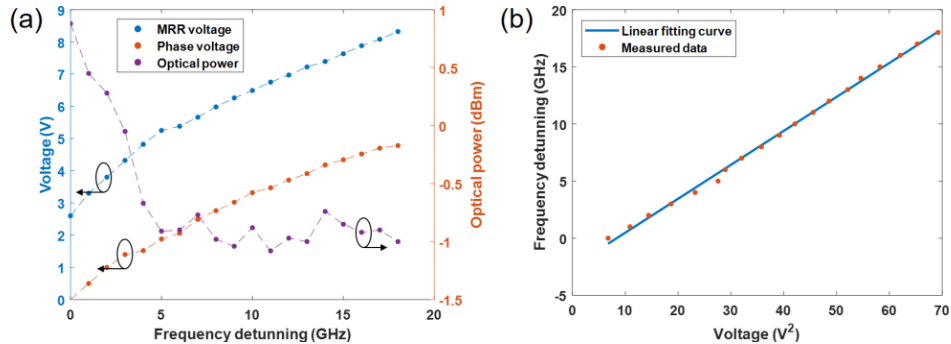

Fig. 5. Characteristics of frequency tuning. (a) MRR and phase tuning voltage with respect to frequency detunning. The measured wavelength band is from 1548.65 nm to 1548.79 nm. Frequency detunning is frequency difference from 1548.65 nm. Starting from 1548 nm, for every 1 GHz detunning, the output power and detunning voltage are scattered in figure. (b) The linearly chirped frequency with respect to MRR detunning power. The square of the MRR voltage represents the thermal power, which is linear with output frequency detunning.

In Fig. 4(a), the 42 GHz continuous frequency tuning band is obtained. For stable linearly chirped FMCW, the wavelength band from 1548.65 nm to 1548.79 nm is selected. When the tuning range exceeds 1548.8 nm, the power gradually decreases. Although it is still a stable single-longitudinal mode operation, there is a risk of mode-hopping during fast frequency modulation. Fig. 4(a) demonstrates the frequency detunning from 1548.65 nm with respect to MRR and phase tuning voltage. As aforementioned two-dimensional tuning method, the MRR is mainly used to tune the output frequency while phase electrode is used to compensation. If the phase compensation is not keeping up with MRR, the laser will be mode hopping. Phase delay $\psi_0$ is complementary to MRR detuning, so that phase voltage $V_p$ has a linear correlation with MRR voltage $V_{mrr}$, as shown in Fig. 5(a). $V_p$ can be written approximately proportional to $V_{mrr}$:

$$V_p = kV_{mrr} \quad (6)$$

where $k$ is the proportional coefficient. Fig. 5(a) also demonstrates that within the tuning band, the output power keeps relatively stable avoiding intensity fluctuation. In Fig. 5(b), it demonstrates that the tuning power of MRR is linear with detuning frequency. The square of voltage represents the thermal power. It is known that the resonant frequency of MRR is linear with thermal power. And the Fig. 5 (b) illustrates that the MRR's resonant frequency is indeed linear with detuning frequency, which experimentally achieved the linear tuning area shown in Fig. 2. If time-dependent voltage $u_{mrr}(t)$ and $u_p(t)$ is applied, the time-dependent frequency modulation $v(t)$ will be obtained. However, the heat propagation with time of thermal electrodes is not instantaneous, resulting in $v(t)$ being nonlinear [27]. Linearly frequency detuning in static state is achieved but not achieved in dynamical state due to the thermal effect. Heater material is about 8 μm above the waveguide, as shown Fig. 3(a). Thinning the PECVD $SiO_2$ cladding layer to achieve faster thermal transmission may be an option.

Here, we employ the iterative learning pre-distortion linearization of the hybrid laser source [2]. It has been proven effective in generating periodic modulated signal [28]. In our experiment, the FMCW laser source is first pass the 20 m-delayed unbalanced MZI. The experimental setup is the same as Fig. 4(a). The PD receives the delayed self-homodyne signal. Then use the Hilbert Transform (HT) to extract the phase $\varphi_b(t)$ of the beat note. Since the time delay is small, the phase can be written as:

$$\varphi_b(t) = 2\pi\tau v(t) \quad (7)$$

Based on the nonlinear phase $\varphi_b(t)$, fitted the desired linear phase $\varphi_d(t)$. Calculate the residual nonlinear frequency difference $v_{res}(t) = [\varphi_b(t) - \varphi_d(t)]/2\pi\tau$. Considering that heat conduction is the first derivative with time, the iterative learning variables need to use derivative of residual difference $\frac{d}{dt}v_{res}(t)$ and $v_{res}(t)$ itself. The n$^{th}$ driver voltage is obtained by iteration:

$$u_{mrr,n}(t) = u_{mrr,n-1}(t) + av_{res}(t) + b\frac{d}{dt}v_{res}(t) \quad (8)$$

where $u_{mrr,n}(t)$ and $u_{mrr,n-1}(t)$ represent the $n^{th}$ and $n-1^{th}$ driving voltage respectively. a and b here represent the iterative learning rate.

Take 500 Hz chirp frequency as an example, by applying initial driving voltage $u_{mrr}(t)$ and $u_p(t)$, chirped frequency is quadratic with time, $1 - R^2$ is calculated to be more than 0.2. By 50 times iterations, the residual frequency is suppressed within 2 MHz as shown in Fig. 6(a), (b). Since the driving signal is a triangular wave, the chirped performance at both turning point is not good. Therefore, define the region of interest (ROI) as the linear chirped frequency region at the center of the chirped band. The ratio of ROI to the entire chirped bandwidth of up-ramp and down-ramp of 500 Hz chirp frequency is 90% and 85% respectively. $v_{rms}$ is the root mean square (RMS) of the residual frequency $v_{res}(t)$. It represents the frequency fluctuation in the chirped band. $R^2$ is the linear regression coefficient representing the linearity of each ramp. $1 - R^2$ is calculated as $3.58 \times 10^{-9}$ and $2.83 \times 10^{-9}$ of up and down ramp respectively, which is 9 order of magnitude improved than initial driving result. Fig. 6(c) is the power spectrum of the beat note generating by beating the FMCW laser source with single frequency reference laser (PPCL-550, Pure Photonics). The chirped bandwidth is 8.63 GHz of the spectrum.

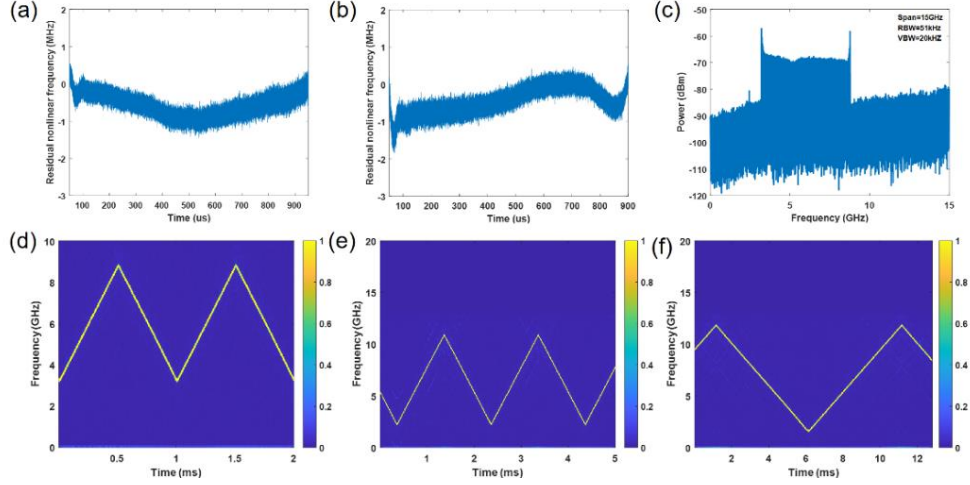

Fig. 6. (a), (b) The residual nonlinear frequency $v_{res}(t)$ of the up-ramp and down-ramp of 500 Hz FMCW signal. (c) Power spectrum of the beat note generated by beating reference laser with FMCW LUT. (d)-(e) The time-frequency diagram of the beat note generated by beating 1k Hz, 500 Hz, 100 Hz FMCW laser source with reference laser, respectively.

**Table 1. Parameters of the generated FMCW signal**

| Chirp Frequency (Hz) | Bandwidth (GHz) | Chirp rate (GHz/ms) | Up-ramp | | | Down-ramp | | |
|---|---|---|---|---|---|---|---|---|
| | | | $1-R^2$ | $v_{rms}$ (MHz) | ROI | $1-R^2$ | $v_{rms}$ (MHz) | ROI |
| 100 | 10.25 | 2.05 | $2.96 \times 10^{-8}$ | 1.1896 | 90% | $2.58 \times 10^{-8}$ | 0.9263 | 90% |
| 500 | 8.63 | 8.63 | $3.58 \times 10^{-9}$ | 0.2573 | 90% | $2.83 \times 10^{-9}$ | 0.3110 | 85% |
| 1000 | 5.56 | 11.12 | $5.29 \times 10^{-9}$ | 0.2181 | 90% | $4.07 \times 10^{-9}$ | 0.2076 | 90% |

In addition to 500 Hz chirp frequency, we also demonstrate the results of 100 Hz and 1 kHz chirp frequency. The time-frequency diagram of the beat note by FMCW laser source with reference laser is shown in Fig. 6(d)-(f). Their parameters are listed in table 1. The frequency bandwidth has 10.25 GHz with 100 Hz sweeping speed which generates a microwave waveform with ultra-large TBWP of $5.1 \times 10^7$ which is of great significance in improving the SNR of long-distance detection. On the other hand, the measurement distance resolution will be degraded by wide linewidth and nonlinear frequency modulation. The residual nonlinearity accumulates as the measurement distance increases, and the 3-dB bandwidth of the received signal will linearly expanded as the distance. If the linearity is perfect, $v_{rms}$ can be ignored, the resolution is only related to the frequency bandwidth $\Delta v$. Resolution becomes the $\Delta z = c/2\Delta v$, which is consistent with commonly used formula. The proposed FMCW laser source utilizing pre-distortion linearization achieving high linear frequency chirped, which has a high resolution in long distance detection without nonlinear compensation.

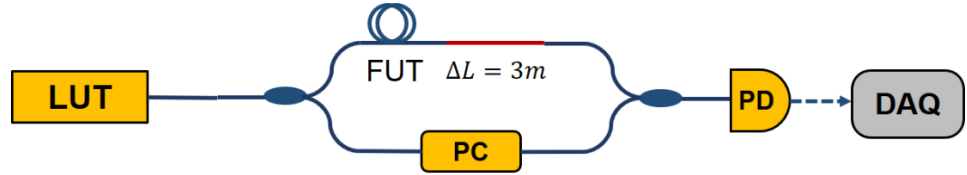

Fig. 7. Experimental setup of FMCW laser source performance evaluation. FUT: Fibre under test, DAQ: data acquisition card.

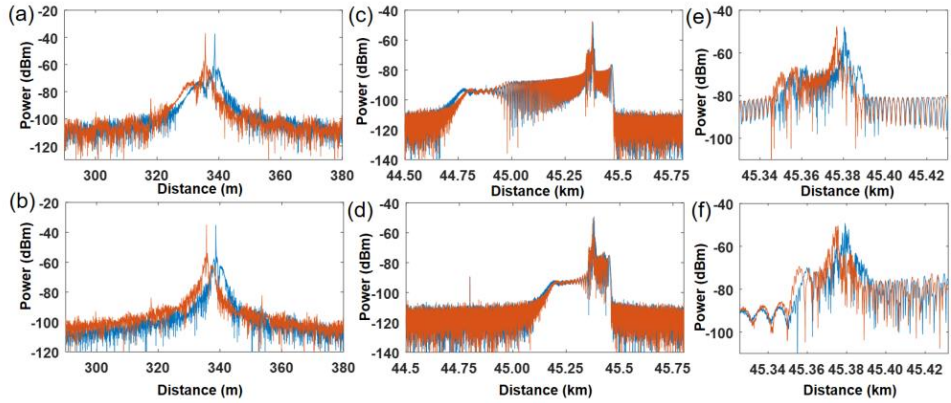

Fig. 8. Spectrum of the beat note by delayed self-homodyne with 1 kHz chirp frequency. Red curve and blue curve represent the result with fibre length difference of 3m. (a), (b) Measure results of up and down ramp with 340 m fibre delay. (c), (d) Measure results of up and down ramp with 45 km fibre delay. (e), (f) The enlarged view at the peak frequency in (c), (d), respectively.

Characteristic of the FMCW laser source is evaluated by measuring the length of the FUT. The experimental setup is shown in Fig. 7. The FUT length is chosen about 350 m and 45 km. In each measurement, a 3 m fibre is connected to the FUT to test the distance resolution. Fig. 8 shows the spectrum of the beat note. Fig. 8(a), (b) are the spectrum of the up and down ramp in one chirped duration period. The red curve in Fig. (a), (b) measures the FUT length is 335.61 m and 335.62 m, respectively. The blue line is the measured result with an extra 3 m fibre in measured arm. The results are 338.65 m and 338.64 m, respectively. So the extra fibre length is calculated as 3.03 m and 3.02 m according to the up and down ramp measured results, respectively. It clearly sperate the two measured spectrum of 3 m length difference. And the 3-dB bandwidth at each peak is about 0.04 m. The theoretical distance resolution at 335.61 m is 0.04 m. Fig. 8(c), (d) shows the measure results of longer FUT. The up and down ramp has the measured distance of FUT is 45376.82 m and 45375.12 m, the distance of FUT with 3 m fibre is measured 45380.43 m 45379.49 m. The extra fibre length is calculated to be 3.61 m and 4.37 m. Compared with the result of hundreds length of FUT, the measured distance and relative resolution length have deteriorated. SNR of 45 km measurement result is 36 dB while it is 70 dB in 400 m measurement. The decrease of SNR is mainly from the spectral tailing caused by frequency nonlinearity. Since the measured length is within the coherent distance, the degradation of SNR due to decoherence is not obvious. The noise floor -103 dBm is quantization noise of DAQ. The results demonstrate the potential application in ultra-long optical fibre sensing and coherent LiDAR. Our results merely use the Fast Fourier Transform (FFT) to process single period detected data without using averaging algorithm and compensation auxiliary MZI. It should be noted that compensation method and algorithms have been developed rapidly in recent years. If more advanced algorithms is applied, the performance of the FMCW laser source will have further improvement [29].

## 5. DISCUSSION

The phase-MRR two-dimensional method for large bandwidth frequency modulation overcomes the bandwidth limitation in self-injection locked state. By thermal tuning, the dynamic range is limited by the thermo-optic (TO) effect. The frequency modulation bandwidth decreases with the increase of modulation speed. In the pre-distortion linearization of FMCW laser source, pre-distortion signals require overload voltage, which limits the bandwidth of the drive signal. The proposed FMCW laser source can generate ultra-large TBWP received waveform thus improves the SNR of the demodulated signal. While it has limitation in fast modulation applications such as 3-D object imaging [30]. The chirped frequency is mainly limited by the thermo-optic (TO) effect so it can by employing carrier-injection-based MRR on silicon-on-insulator (SOI) platform to improve the chirped frequency [31].

## 6. CONCLUSIONS

A hybrid FMCW laser based on self-injection is proposed. The static tuning range is up to 42 GHz with 49.86 Hz intrinsic linewidth. Dynamical FMCW by two-dimensional thermal tuning achieve a large TBWP of $5.1 \times 10^7$. The ultra-long length fibre length is measured which demonstrate potential for application in FMCW LiDAR and fibre sensing.

**Funding.** National Key R&D Program of China under Grant (2018YFB2201802); National Natural Science Foundation of China (NSFC) (61771285);

**Disclosures.** The authors declare no conflicts of interest.